\documentclass[aps,prd,twocolumn,nofootinbib]{revtex4-1}
\usepackage{epsfig}
\usepackage[colorlinks,linkcolor=blue,anchorcolor=blue,citecolor=blue,urlcolor=blue,breaklinks=true]{hyperref}
\usepackage{graphics}
\usepackage{color}
\usepackage{graphicx}
\usepackage{amsmath}
\usepackage{amsfonts}
\usepackage{upgreek}
\usepackage{slashed}
\usepackage{color}

\begin{document}
\author{Wenkai Fan$^{1}$, Xiaofeng Luo$^{2}$, and  Hong-Shi Zong$^{3,4,5}$}
\address{$^{1}$ Kuang Yaming Honors School, Nanjing University, Nanjing 210093, China}
\address{$^{2}$ Key Laboratory of Quark \& Lepton Physics (MOE) and Institute of Particle Physics, Central China Normal University, Wuhan 430079, China}
\address{$^{3}$ Department of Physics, Nanjing University, Nanjing 210093, China}
\address{$^{4}$ Joint Center for Particle, Nuclear Physics and Cosmology, Nanjing 210093, China}
\address{$^{5}$ State Key Laboratory of Theoretical Physics, Institute of Theoretical Physics, CAS, Beijing 100190, China}

\title{Susceptibilities of Conserved Charges within a Modified Nambu-Jona-Lasinio Model}
\begin{abstract}
We modified the original Nambu-Jona-Lasinio model by considering the feedback of quark propagator on the coupling. By doing this, we found that the chiral condensate at finite temperature calculated from Lattice QCD can be well fitted by this modified NJL model. With this model, we calculate the susceptibilities of the baryon, charge and strangeness as a function of temperature and baryon chemical potential in the QCD phase diagram. The baryon number susceptibilities are found to be of the greatest magnitude, whereas the strangeness susceptibilities have the smallest divergence dominating area due to the large strange quark mass. Furthermore, the strange quark¡¯s chemical potential is found to be of minor influence on the susceptibilities. Finally, we studied the energy dependence of these susceptibilities along the chemical freeze-out line obtained from heavy-ion collisions experiments.
\end{abstract}


\maketitle

\section{Introduction}
Exploring the phase structure of strongly interacting nuclear matter is one of the main goals of heavy-ion collision experiments. Lattice QCD calculations show that at small baryon chemical potential and high temperature, the transition from the hadron gas phase to the Quark Gluon Plasma (QGP) phase is a smooth crossover \cite{fodor2002new}, whereas a first--order phase transition is expected at high baryon chemical region \cite{masayuki1989chiral,halasz1998phase,stephanov2006qcd,ejiri2008canonical,Lu:2015naa,Lu:2016uwy}. The end point of the first--order phase boundary towards the crossover region is called the QCD critical end point (CEP). It has long been predicted that the fluctuations and correlations are sensitive to the phase transition and can be used to study the phase structure of strongly interacting nuclear matter.  The experimental measurements of the fluctuations of conserved quantities have been performed in the beam energy scan (BES) program by the STAR and PHENIX experiments at the Relativistic Heavy-Ion Collider (RHIC). Interestingly, the STAR experiment observed a non-monotonic energy dependence of the fourth order ($\kappa\sigma^{2}$) net-proton fluctuations in the most central Au+Au collisions. Furthermore, this non-monotonic behavior cannot be described by various transport models \cite{xu2016cumulants,He2016296,Luo201675}. To investigate the contribution of the critical point physics to the conserved charges fluctuations and their energy dependence behavior from theoretical calculations, we have calculated the various fluctuations along the freeze-out line in the QCD phase diagram with a modified Nambu-Jona-Lasinio (NJL) model \cite{nambu1961dynamical,klevansky1992nambu}. Previous work has studied these quantities up to fourth order \cite{hatta2003universality,asakawa2009third,chen2015baryon,chen2016robust}, or made use of the Polyakov-loop improved NJL model \cite{fu2010fluctuations,Cui:2013aba}. Other effective model like Polyakov-Quark-Meson (PQM) model \cite{friman2011fluctuations,skokov2012non} give similar qualitative behavior of the fluctuations like in the NJL model. Discussion on the signs of these quantities using a more general approach in Refs.~\cite{stephanov2009non,stephanov2011sign} also agree with the NJL model calculation.

In our work, we use a modified three flavor NJL model, with the four-point coupling being dependent on the quark condensate, inspired by the Operator Product Expansion (OPE) method \cite{shi2016continuum}. Due to the sign problem of lattice simulation at finite chemical potential, lattice method are limited to low baryon chemical potential and finite temperature right now and conventional NJL model can not match lattice result at zero chemical potential and finite temperature. However, just as we will show below, with a coupling strength depending on the quark condensate, we are able to reproduce lattice result at finite temperature and $\mu_B=0$. This makes an extension to finite chemical potential more reliable. From experiment data \cite{das2015chemical}, we know the chemical potential of $u,d$ quarks are almost the same, so we set them to be equal throughout the calculation. The chemical potential of the strange quark is smaller, but due to the large mass of $s$ quark, it does not vary the phase diagram much, thus having small influence on the susceptibilities. We first set the chemical potential of the $s$ quark to be the same as $u,d$ quarks for simplicity. A discussion on the effect of the strange quark's chemical potential is given in Sec.~\ref{sec:mus_compare} and we found the choice of $\mu_s$ does not change our result much, as expected. We also calculated the susceptibilities of a non-interacting quark gas to see the influence of critical behavior on the susceptibilities, especially the sign of these susceptibilities and their energy dependence. Our model calculation is in agreement with previous more general argue of the signs of these susceptibilities \cite{stephanov2009non,asakawa2009third}. Throughout our calculation, we assume that the fire-ball is near thermal equilibrium at freeze-out, though finite size effect, critical slowing of dynamics when the fire-ball passes the CEP \cite{berdnikov2000slowing,athanasiou2010using}, changes in expansion dynamics and interactions that produce variations in particle spectra and acceptance \cite{koch2010hadronic} might blur the signal of criticality and must be controlled for.

This paper is organized as follows: In the second section ,we introduce the modified NJL model. In section three we give the definition of the various quark number susceptibilities. Then we calculate two quantities that can be compared to experiment measurement in the next two sections. The energy dependence of these two quantities are given in section five. The effect of the strange quark chemical potential is discussed in section six. In section seven, we compare our results with previous work and HRG model calculation at zero chemical potential. Finally, we summarize our work and give an outlook in future experiment.
\begin{figure*}[htbp]
  \centering
  \includegraphics[width=0.74\textwidth]{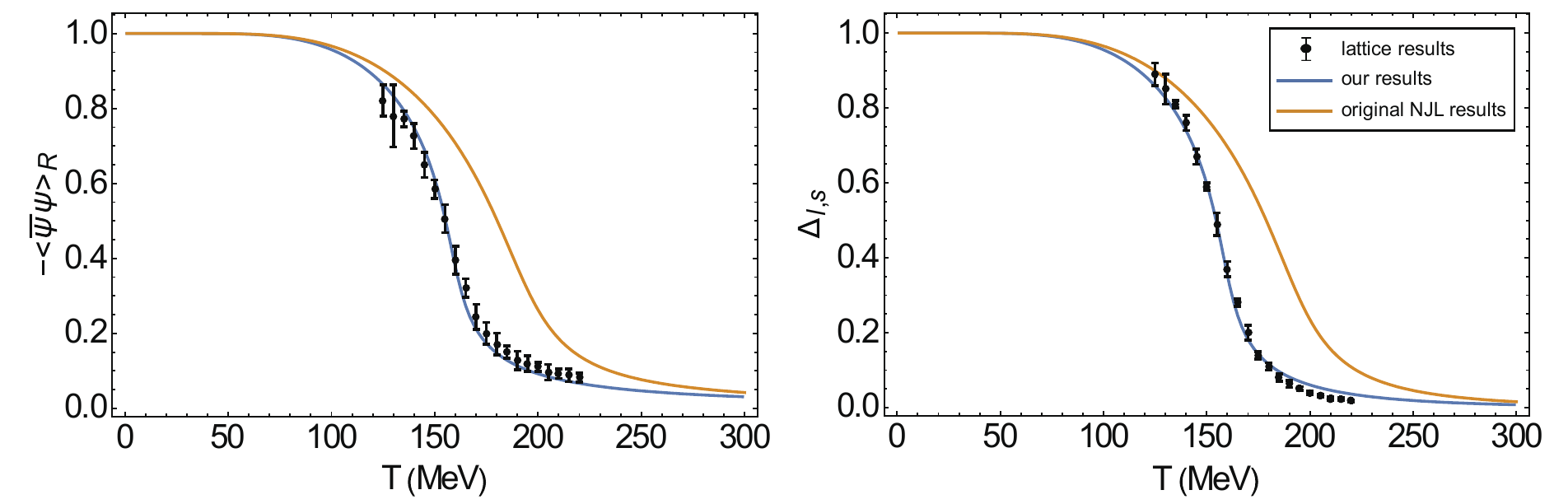}\\
  \caption{(a) Normalized light quark ($u,d$) condensate versus $T$, compared to lattice result from Ref.~\cite{borsanyi2010there} (b) $\Delta_{l,s}$ (a linear combination of $\langle\overline{u}u\rangle$ and $\langle\overline{s} s\rangle$) versus $T$}\label{fig:lattice_compare}
\end{figure*}
\begin{figure*}[htbp]
  \centering
  \includegraphics[width=0.72\textwidth]{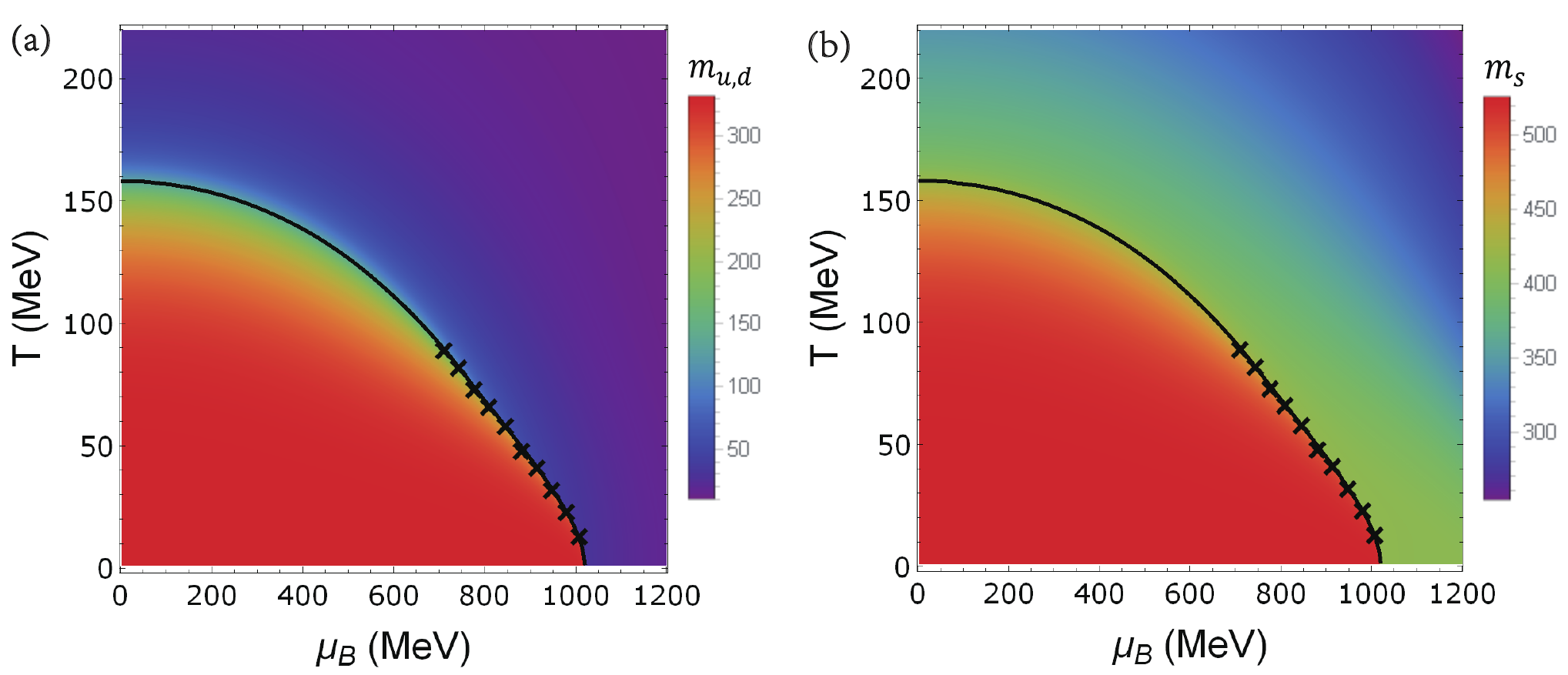}\\
  \caption{The phase diagram of quark masses. The solid line is the crossover line and the crosses make up the first order phase transition line (a) Up quark mass $m_u$ and Down quark mass $m_d$ (b) Strange quark mass $m_s$}\label{fig:quark_mass_all}
\end{figure*}
\section{Modification of the NJL model}\label{2}
The commonly used lagrangian density of the NJL model with $2+1$ flavor is:
\begin{equation}\label{equ:original_lagrangian}
\begin{aligned}
\mathcal{L}=\overline{\psi}(i\slashed \partial-m)\psi+G[(\overline{\psi}\lambda_i\psi)^2+(\overline{\psi}i \gamma_5 \lambda_i\psi)^2]\\
-K({det[\overline{\psi}(1+\gamma_5)\psi]+det[\overline{\psi}(1-\gamma_5)\psi]})
\end{aligned}
\end{equation}

After doing mean--field approximation of the Lagrangian density in Eq.~\eqref{equ:original_lagrangian}, we have the following gap equation and expressions for quark condensate and number density:
\begin{equation}\label{equ:0th order}
\left\{
\begin{aligned}
&m_i=m_{i0}-4G \langle\overline{q}_i q_i\rangle+2K \langle\overline{q}_m q_m\rangle \langle\overline{q}_n q_n\rangle (i\neq m \neq n)\\
&\langle\overline{q}_i q_i\rangle=-m_i F(m_i,\mu_i)\\
&\langle q_i^{\dag}q_i\rangle=H(m_i,\mu_i)
\end{aligned}
\right.
\end{equation}
where $\langle\Theta\rangle=\frac{Tr (\Theta e^{-\beta(\mathcal{H}-\mu_i \mathcal{N}_i)})}{Tr( e^{-\beta(\mathcal{H}-\mu_i\mathcal{N}_i)})}$ being the grand canonical ensemble average, and $i=u,d,s$. And we define:
\begin{equation}\label{eq:FandH}
\begin{aligned}
&F(m,\mu)=\frac{N_c}{\pi^2}\int^{\Lambda}dp\frac{p^2}{E_p}(1-f^{-}(m,\mu)-f^{+}(m,\mu))\\
&H(m,\mu)=\frac{N_c}{\pi^2}\int^{\Lambda}dp p^2(f^{-}(m,\mu)-f^{+}(m,\mu))
\end{aligned}
\end{equation}
where $f^{\pm}(m,\mu)=1/(1+E^{\beta(E\pm\mu)})$, and $N_c=3$.

The four-point coupling $G$ has a physical meaning of being an effective gluon propagator. If we take the quark propagator's feed back on the gluon self-energy into consideration \cite{steele1989quark,jiang2012wigner,Cui:2013tva}, the coupling strength is replaced by:
\begin{equation}\label{G1G2}
\begin{aligned}
 G=G_1+G_2(\langle\overline{u}u\rangle+\langle\overline{d}d\rangle)+G_3\langle\overline{s}s\rangle
\end{aligned}
\end{equation}

In this paper we restrict our discussion within the $G_2=G_3$ case for simplicity. Also, the $6$-point coupling constant $K$ is kept constant since the change of it will have much smaller effect than that of $G$. We can fix these parameters at zero temperature and chemical potential following standard procedure and we adopt the parameter set used in Ref.~\cite{hatsuda1987effects}. The ratio of $G_1$ to $G_2(G_3)$ is further determined by fitting the critical temperature $T_c$ to lattice results at finite temperature \cite{borsanyi2010there}.
\begin{table}[htbp] \centering%
\begin{tabular}{|c|c|c|}
\hline
$m_u(MeV)$           &$m_s(MeV)$             & $\Lambda(MeV)$      \\
\hline
   $5$               & $136$                 & $631$               \\
\hline
$G_1(MeV^{-2})$      & $G_2(G_3)(MeV^{-5})$  &   $K(MeV^{-5})$     \\
\hline
$3.74\times10^{-6}$  & $-1.74\times10^{-14}$ &$9.29\times10^{-14}$ \\
\hline
\end{tabular}
\caption{Parameter set used in our work for the NJL model}\label{tab1}
\end{table}
With only one more parameter, we are able to fit the $\langle\overline{u}u\rangle$ and $\langle\overline{s} s\rangle$ at finite temperature to lattice results quite well. The critical temperature at zero chemical potential is about $158MeV$. One should notice that both the NJL model and the lattice simulation we referred to use physical quark masses and other parameters fitted to physical meson properties, which is important since the behavior of both theories depends strongly on the parameter sets they adopt. Previous NJL model calculation gives a $T_c$ at about $170MeV$ \cite{chen2016robust} and the PNJL model gives a $T_c$ of $200MeV$ \cite{fukushima2008phase} with the same parameter set in Table.~\ref{tab1}.

In real experiments, although $\mu_u\approx\mu_d\approx\frac{1}{3}\mu_B$, $\mu_s$ varies at different collision energy from about $\frac{1}{5}\mu_B$ to $\frac{1}{3}\mu_B$ \cite{das2015chemical,Chatterjee2015Freeze}. Since the location of the CEP is weakly dependent on the choice of $\mu_s$ (see Sec.~\ref{sec:mus_compare}), we first assume equal chemical potential for the three quarks. The resulting phase diagram of the quark masses are plotted in Fig.~\ref{fig:quark_mass_all}. The quark masses remain almost constant at low chemical potential and temperature, but varies drastically along a certain ``band'' when $T$ and $\mu_B$ increase, which indicates a phase transition. At high baryon chemical potential or temperature, up and down quarks become small while the strange quark remains quite massive. The CEP is located at $(\mu_B,T)=(711MeV,90MeV)$. Because of the discontinuity of the quark mass, the susceptibilities should have a divergent behavior near the CEP.

\section{Quark number susceptibility derivation}
As the linear response of the physical system to some external field, susceptibility is often measured to study the properties of the related system. Therefore the studies of various susceptibilities are very important on the theoretical side, which are widely used to study the phase transitions of strongly interacting matter \cite{Cui:2015xta}. The various susceptibilities are defined as:
\begin{equation}
\begin{aligned}
&\frac{\partial\langle q_i^{\dag}q_i\rangle}{\partial \mu_j}=\chi_{i,j},\frac{\partial^2\langle q_i^{\dag}q_i\rangle}{\partial \mu_j\partial \mu_k}=\chi_{i,jk},\frac{\partial^3\langle q_i^{\dag}q_i\rangle}{\partial \mu_j\partial \mu_k\partial \mu_p}=\chi_{i,jkp}
\end{aligned}
\end{equation}
We calculate these susceptibilities by explicitly taking derivatives of the gap equations in Eq.~\ref{equ:0th order}. This involves much symbolic manipulation of the various derived equations but has the virtue that it is accurate compared to approximating these susceptibilities by finite difference of the quark densities. Furthermore, we change the base from $\{u,d,s\}$ at quark level to the conserved charges $\{B,Q,S\}$ by using:
\begin{equation}\label{equ:basechange}
\left\{
\begin{aligned}
&\mu_u=\frac{1}{3}(\mu_B+2\mu_Q)\\
&\mu_d=\frac{1}{3}(\mu_B-\mu_Q)\\
&\mu_s=\frac{1}{3}(\mu_B-\mu_Q-3\mu_S)
\end{aligned}
\right.
\end{equation}

Then the various susceptibilities can be expressed in the basis of $\{B,Q,S\}$. For example:
\begin{equation}\label{equ:susbasechange}
\begin{aligned}
 \chi_{B}^{(n)}&=\frac{1}{3^n}\sum_{\{i,j,k...\}=u,d,s}\chi_{i,j,k...}\\
  \chi_{Q}^{(n)}&=\frac{1}{3^n}\sum_{\{i,j,k...\}=u,d,s}2^p(-1)^q(-1)^r\chi_{i,j,k...}\\
  (p&,q,r\ equals\ the\ number\ of\ u,d,s\\
   in &\ \{i,j,k...\}\ respectively)\\
  \chi_{S}^{(n)}&=(-1)^{n}\chi_{s,s,s...}\\
  \end{aligned}
\end{equation}
The whole process is applicable to effective models after doing mean--field approximation. Only Eq.~\eqref{equ:0th order} needs to be modified according to the specific model.
\begin{figure*}[htbp]
  \centering
  \includegraphics[width=0.96\textwidth]{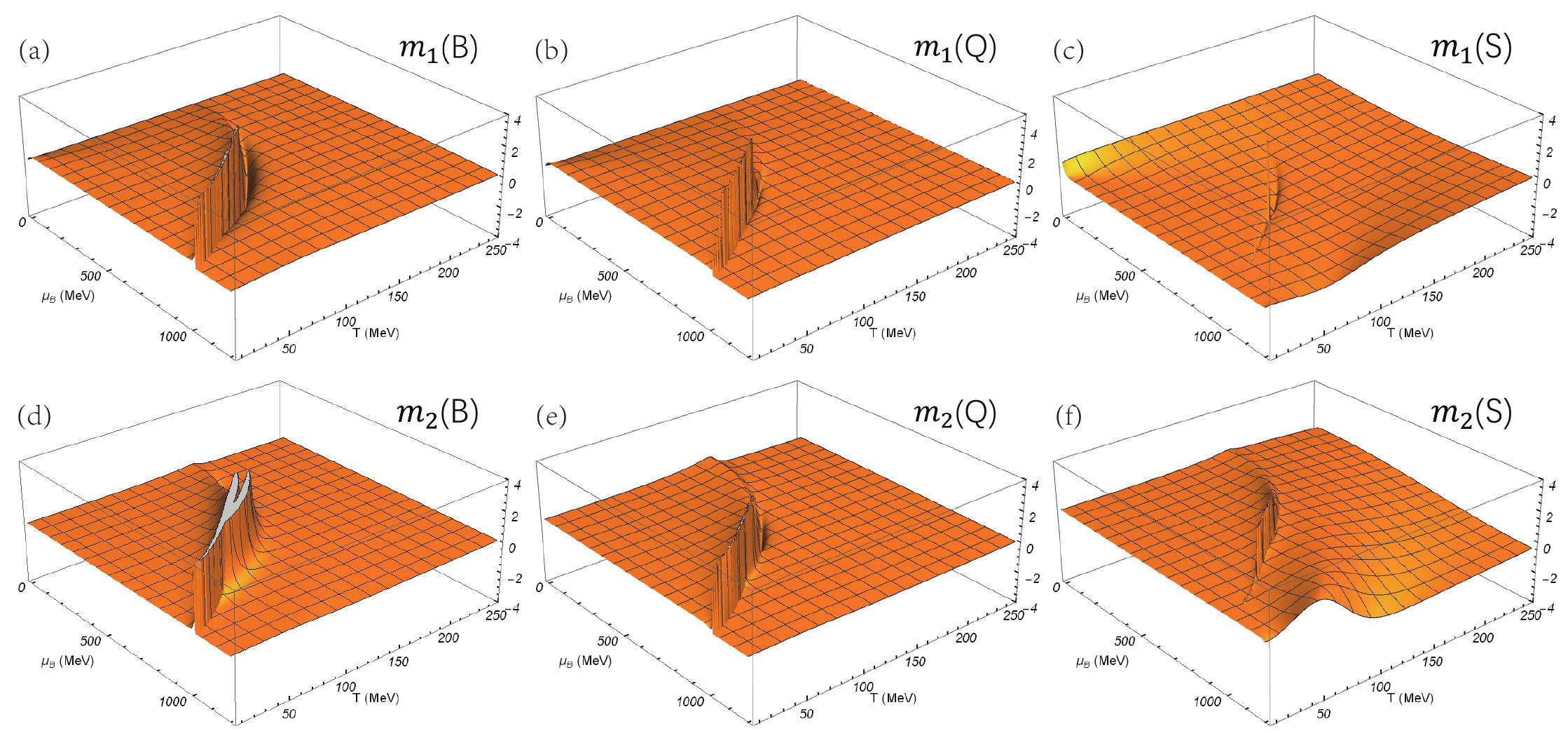}\\
  \caption{$m_1$ and $m_2$ of baryon number, charge number and strangeness (a) $m_1(B)$ (b) $m_1(Q)$ (c) $m_1(S)$ (d) $m_2(B)$ (e) $m_2(Q)$ (f) $m_2(S)$}\label{fig:m_all}
\end{figure*}
\begin{figure*}[htbp]
  \centering
  \includegraphics[width=0.96\textwidth]{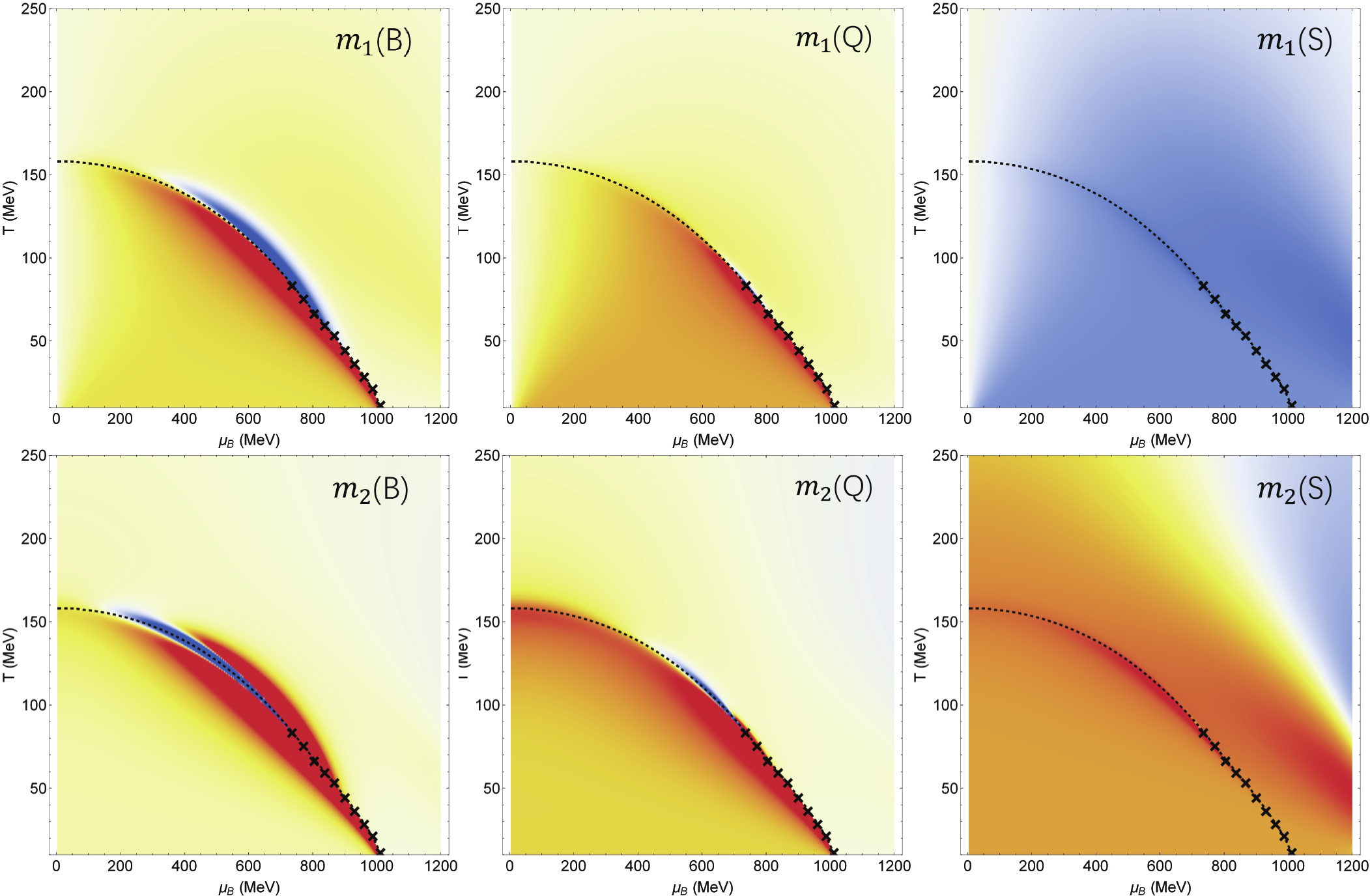}\\
  \caption{Sign of $m_1$ and $m_2$ of baryon number, charge number and strangeness. Red region represents positive value while blue zone represents negative value. The dashed line is the crossover line while the crosses represents the first--order phase transition curve (a) $m_1(B)$ (b) $m_1(Q)$ (c) $m_1(S)$ (d) $m_2(B)$ (e) $m_2(Q)$ (f) $m_2(S)$}\label{fig:m2d_all}
\end{figure*}

\section{moments of baryon, charge and strange number}
In order to relate our calculation with experiments and other model calculation, we consider the following two ratios defined as:
\begin{equation}\label{}
  m_1(x)=\frac{T\chi_{x}^{(3)}}{\chi_{x}^{(2)}}, m_2(x)=\frac{T^2\chi_{x}^{(4)}}{\chi_{x}^{(2)}}
\end{equation}
where $x=B, Q, S$. These two ratios are then independent of the volume of the system. The signs of $m_1$ and $m_2$ are shown in Fig.~\ref{fig:m2d_all}. Red regions are of positive value, and blue regions are of negative value. The yellow regions represent values very close to $0$. $m_1(B)$ and $m_1(Q)$ show a change of sign across the crossover curve. This is also observed in Ref.~\cite{asakawa2009third}, where $m_1(B)$ and $m_1(Q)$ take negative values outside the phase boundary and the area where $m_1(B)$ takes negative value is much larger than that of the $m_1(Q)$. $m_2(B)$ and $m_2(Q)$ have two negative areas and one positive area which is easy to understand since $\chi_{x}^{(4)}$ is the derivative of $\chi_{x}^{(3)}$. However, the baryon number moments' signal are stronger than that of the charge number and this difference will become more significant if one continues measuring higher moments. $m_1(S)$ and $m_2(S)$ only have large values near the the first--order phase transition curve, but remains close to $0$ in all other area.
\begin{figure}[htbp]
  \centering
  \includegraphics[width=0.48\textwidth]{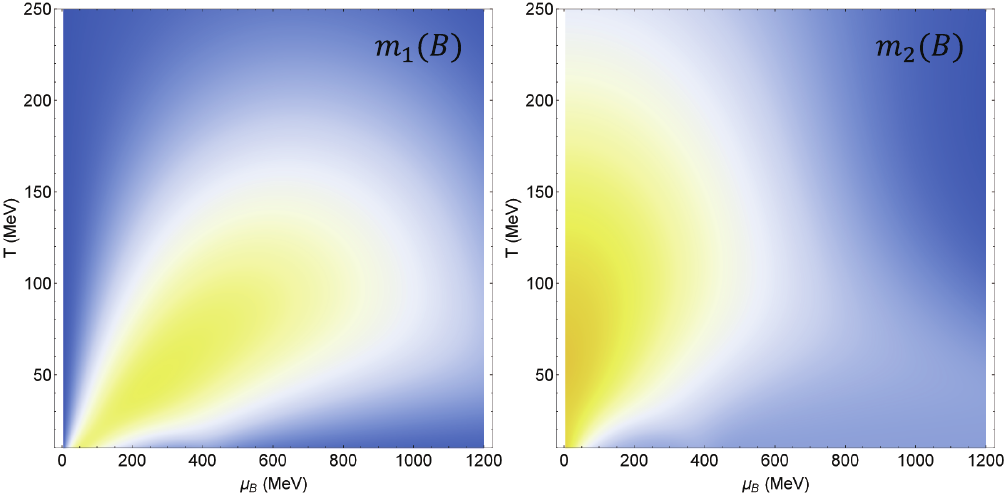}\\
  \caption{Sign of $m_1(B)$ and $m_2(B)$ of the free quark gas case. Both $m_1(B)$ and $m_2(B)$ are close to $0$ and slowly varies with different temperature and chemical potential.}\label{fig:free_m1m2B}
\end{figure}

We also consider an extreme case, $G,K\rightarrow0$, a non-interacting free quark gas case. Thus $m_i=m_{i0}$ at any temperature and chemical potential. The cut-off momentum $\Lambda$ is set to be unchanged since it may not alter the qualitative behavior of $m_1$ and $m_2$. The chemical potentials of the three quarks are again set equal for simplicity. This model is in no case realistic but will give $m_1(x)$ and $m_2(x)$ that are purely contributed by the $\mu$ derivative of the $H(m,\mu)$ function (the quark number density). Thus we can identify how interactions between quarks influence the susceptibilities at finite temperature and baryon chemical potential. We will compare their collision energy dependence with the NJL model results in the next section.

\section{Moments versus collision energy along three hypothetical freeze--out lines}\label{sec:compare_theo}
To obtain the energy dependence of the conserved quantities fluctuations, we need to know the position of the freeze-out line in the QCD phase diagram\cite{cleymans2006status}. Since in RHIC experiments, chemical potential and temperature of the fireball have a distribution along the freeze-out line \cite{Chatterjee2015Freeze,begun2016updates}, we search the region near the phase transition to find whether the theoretical results are consistent with experiments. Owing to the good agreement with lattice result at zero chemical potential and finite temperature, the experimental freeze-out curve \cite{begun2016updates} is close to the theoretical crossover line at high collision energy. Although this model is expected to be not accurate, we found our calculation results are with similar energy dependence trend as observed in the experimental data. At high collision energy $\sqrt{s}$, the trend of $m_1(B), m_2(B)$ are consistent between our calculations and data. Same oscillation patterns are found for $m_1(B), m_2(B)$ at low $\sqrt{s}$. However, at very low collision energy (less than a few $GeV$), $m_1(B), m_2(B)$ may behave differently, seen from our theoretical calculation. It should be meaningful to measure $m_1(B), m_2(B)$ below a few $GeV$ of $\sqrt{s}$ in future RHIC experiments.

\begin{figure}[htbp]
  \centering
  \includegraphics[width=0.42\textwidth]{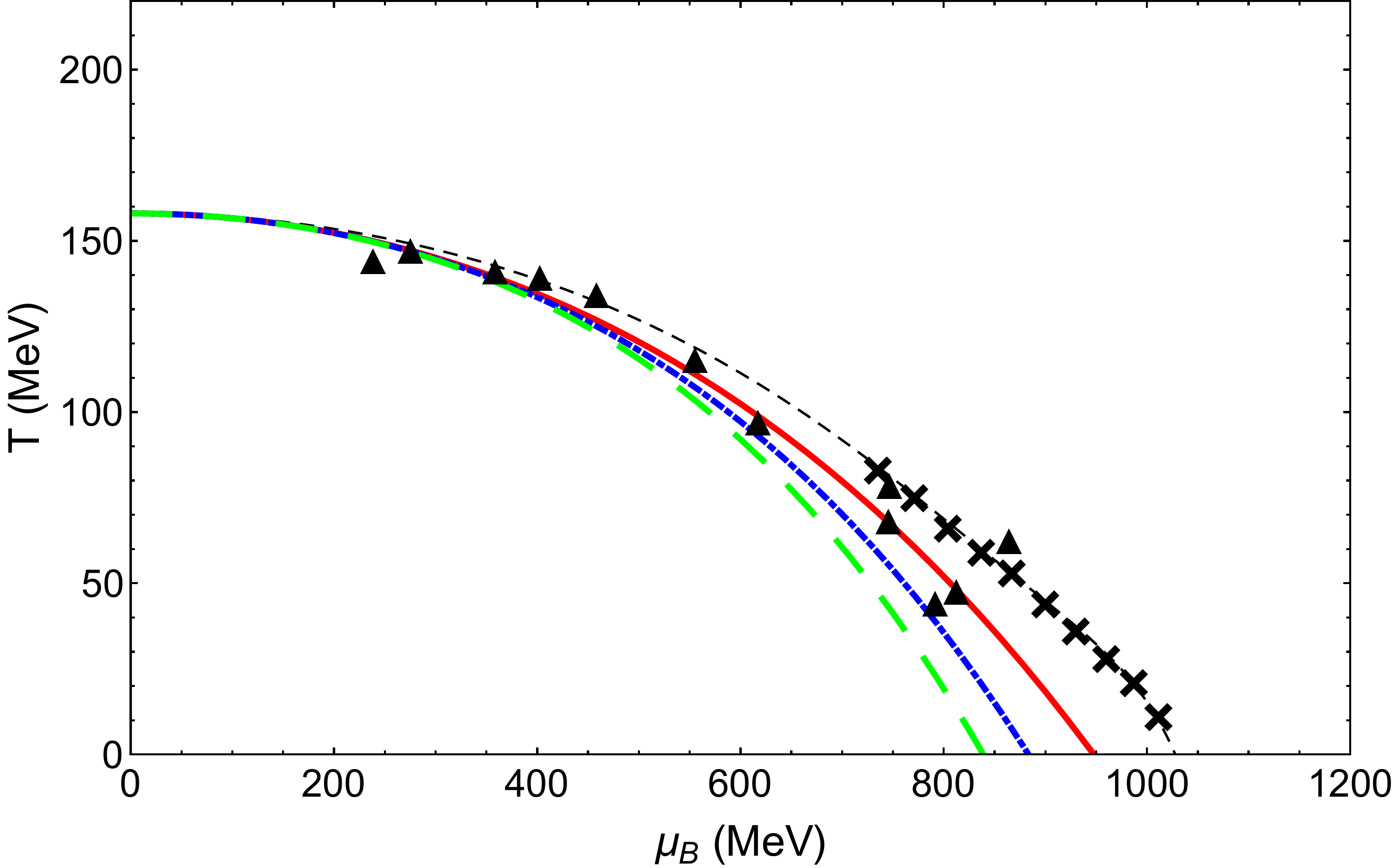}\\
  \caption{Three possible freeze-out curves. The dashed line is the crossover line. Crosses indicate the curve of the first--order phase transition. The triangles are experimental data taken from Ref.~\cite{begun2016updates}. The red solid freeze-out curve is fitted to experimental data and the other two freeze-out curves differ from it by a small amount (see Eq.~\ref{eq:freezeout})}\label{fig:freezeout}
\end{figure}

In real experiments, we can tune the temperature and baryon chemical potential of the system by varying the collision energy. This correspondence is reflected on the chemical freeze-out curve. However, for different collision centrality (or some other parameters), the freeze-out curve may shift a little in the phase diagram. In Fig.~\ref{fig:freezeout}, the dashed line is the crossover line along with the first--order phase transition line. The three colored lines are three hypothetical freeze--out lines with one of them fitted to recent experimental data taken from \cite{begun2016updates}. The formula for these three curves are:
\begin{equation}\label{eq:freezeout}
  T(\mu_B)=a-b\mu_B^2-c\mu_B^4
\end{equation}
where $a=0.158GeV$, $b=0.14GeV^{-1}$, and $c=0.04$ (solid), $0.08$ (dot-dashed), $0.12$ (dashed)$GeV^{-3}$. Since we don't know the exact correspondence between experiment parameters and the location on the phase diagram, we set these three curves to search for a rather large region.

Another formula relating collision energy and baryon chemical potential is \cite{begun2016updates}:
\begin{equation}\label{equ:muBs}
  \mu_B(\sqrt{s})=\frac{1.477GeV}{1+0.343GeV^{-1}\sqrt{s}}
\end{equation}
\begin{figure*}[htbp]
  \centering
  \includegraphics[width=0.78\textwidth]{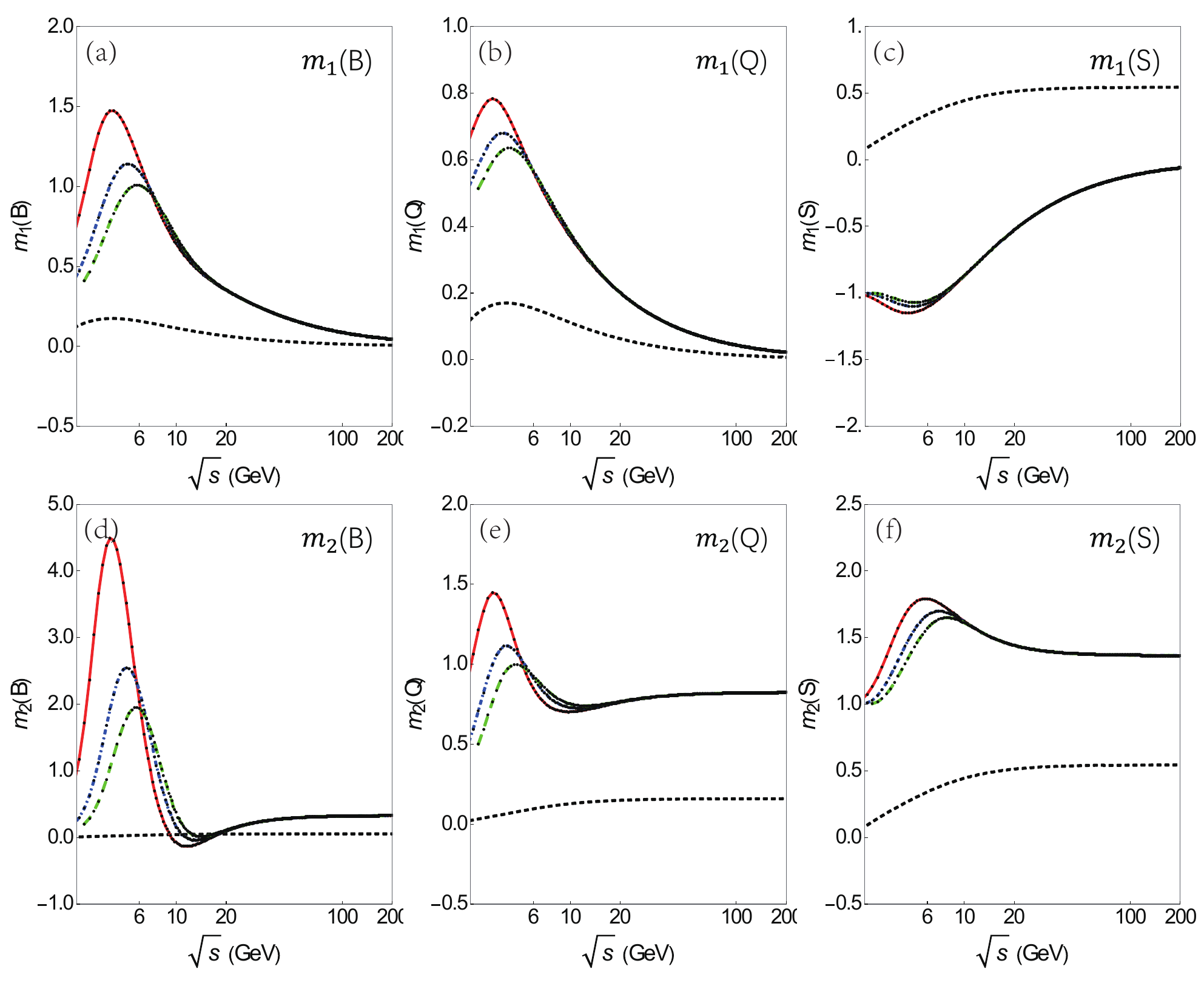}\\
  \caption{Moments versus collision energy along the three hypothetical freeze--out lines, the colored lines are identified in Fig.~\ref{fig:freezeout}, the black dashed lines are results from a free quark gas model. (a) $m_1(B)$ (b) $m_1(Q)$ (c) $m_1(S)$ (d) $m_2(B)$ (e) $m_2(Q)$ (f) $m_2(S)$}\label{fig:m1m2all}
\end{figure*}
%

With the freeze--out curve and Eq.~\eqref{equ:muBs}, we plot the $m_1,m_2$ of $B,Q,S$ versus collision energy $\sqrt{s}$ in Fig.~\ref{fig:m1m2all}. The three lines are identified by color with those in Fig.~\ref{fig:freezeout}. The black dashed lines are the results from the free quark gas model. At low collision energy, the NJL model predicts oscillatory signal of the moments while for the free gas case all moments are close to $0$. Also, we can infer that $m_2(B)$ should be a better probe towards the critical behavior of the susceptibilities since it is larger in magnitude, having one more change in sign of the derivative with respect to collision energy than $m_1(B)$, and is much more different from the weakly interacting quark gas prediction. Also, the magnitude is much smaller (but closer to experiment) this time. The behavior of these two quantities $m_1(B)$ and $m_2(B)$ at $\sqrt{s}$ below a few $GeV$ where experiments have not covered yet are of great importance since some other models predict opposite slope of these two quantities compared to the NJL prediction \cite{fukushima2015hadron,mukherjee2016high}.

\section{Effect of the strange quark chemical potential}\label{sec:mus_compare}
In experiment, the strange quark chemical potential varies at different collision energy from about $\frac{1}{5}\mu_B$ to $\frac{1}{3}\mu_B$ \cite{das2015chemical,Chatterjee2015Freeze}, i.e. about $\mu_s=0.6\mu_u \sim \mu_u$. We argued in previous section that the chemical potential of the strange quark will have small effect on the phase diagram and susceptibilities due to the large mass of the strange quark. In this section we consider two extreme cases: $\mu_s=0$ and $\mu_s=\mu_B/3$. We first show comparison of the crossover line and first-order transition line between the two cases. Then the plot of $m_1(B),m_2(B)-\sqrt{s}$ between the two cases are given. The crossover line and the first-order phase transition boundary between the two cases are very close to each other, the location of CEP differs by about $10MeV$ in $T$ and $30MeV$ in $\mu_B$.
\begin{figure}[htbp]
  \centering
  \includegraphics[width=0.36\textwidth]{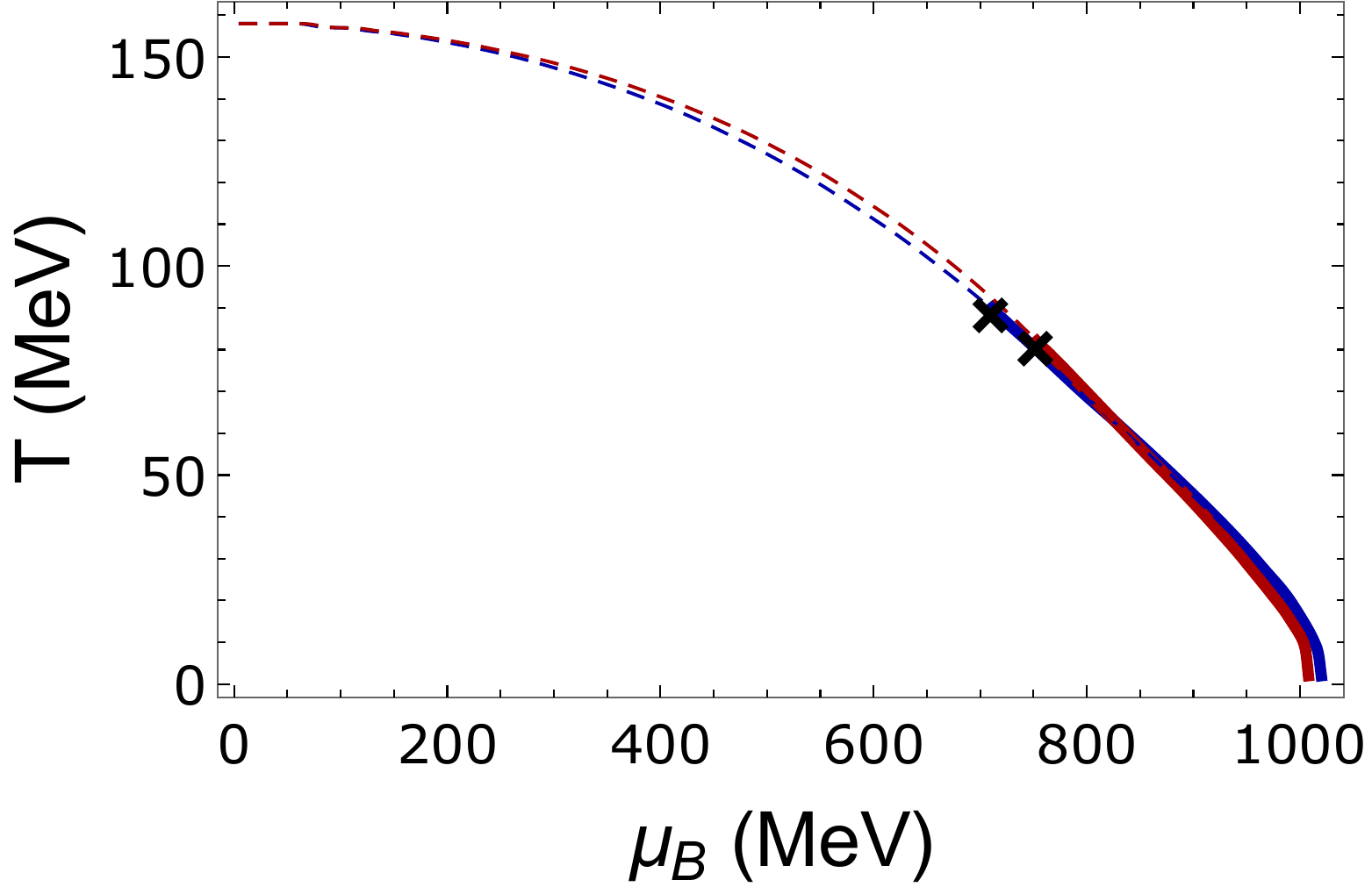}\\
  \caption{Crossover line and first-order phase transition boundary of the $\mu_s=0$ and $\mu_s=\mu_B/3$ cases. The crosses are the two CEP points}\label{fig:phase_dia_compare}
\end{figure}

\begin{figure}[htbp]
  \centering
  \includegraphics[width=0.5\textwidth]{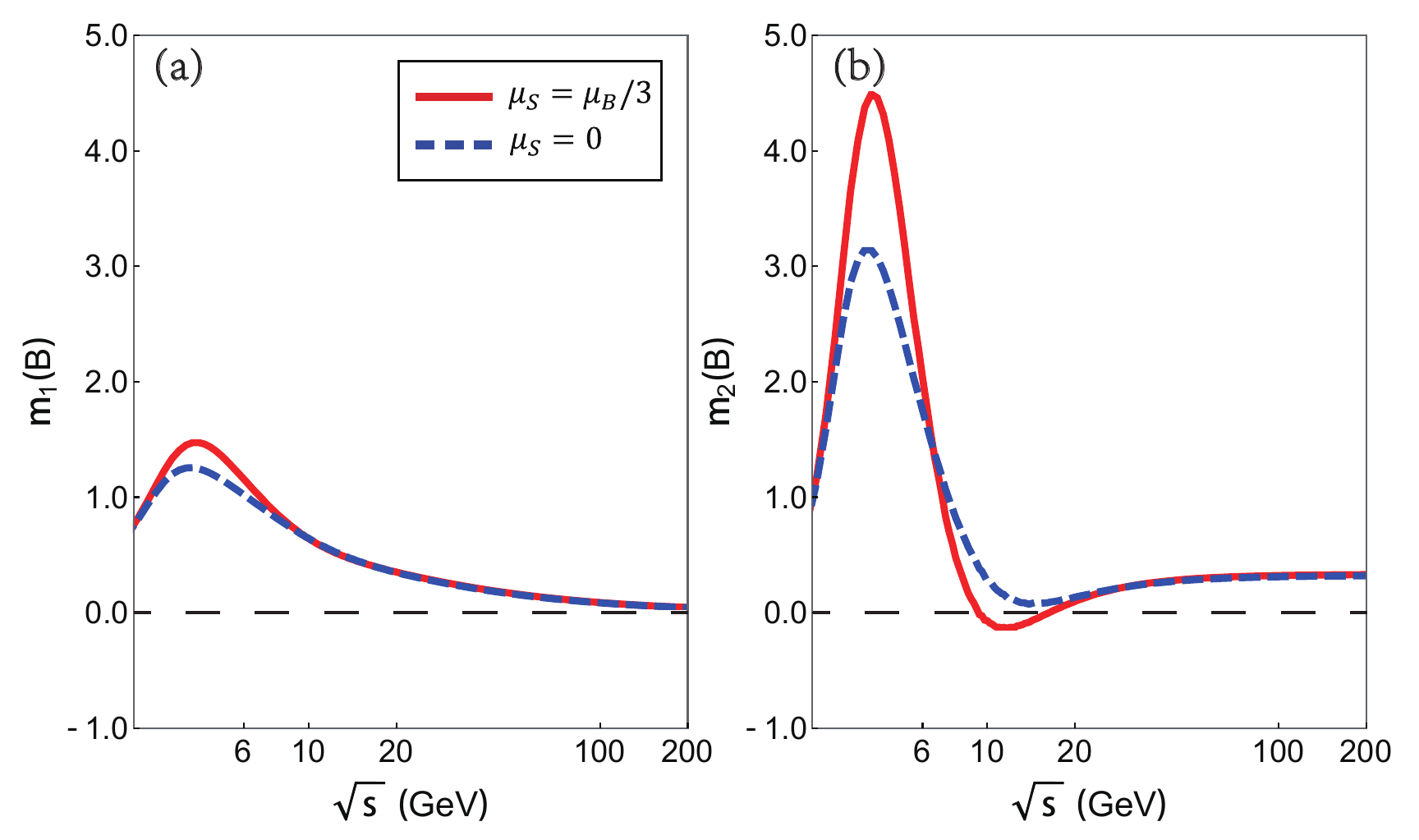}\\
  \caption{(a) $m_1(B)-\sqrt{s}$ (b) $m_2(B)-\sqrt{s}$ for the two cases along the red solid freeze-out line (fitted to experiment data) in Fig.~\ref{fig:freezeout}.}\label{fig:m1m2_compare}
\end{figure}

The various susceptibilities are found to be of no big differences between the two cases. We plot the $m_1(B),m_2(B)-\sqrt{s}$ for illustration. As seen in Fig.~\ref{fig:m1m2_compare}, smaller $\mu_s$ will decrease both the magnitude of $m_1(B)$ and $m_2(B)$.

\section{Comparison with the HRG model at zero chemical potential}\label{sec:HRG_compare}
The Hadron Resonance Gas (HRG) model describes the fireball as a gas of uncorrelated hadrons \cite{huovinen2010qcd,borsanyi2015fluctuations}. It predicts $m_1(B)=0$ and $m_2(B)=1$ at finite temperature and $\mu_B=0$. This can be easily seen from the pressure of the HRG model below:
\begin{equation}\label{equ:HRG pressure}
  \frac{p}{T^4}=\sum_{i=1} K(T,m_i) cosh(B_i\hat{\mu}_B+Q_i\hat{\mu}_Q+S_i\hat{\mu}_S)
\end{equation}
where the sum runs over the species of hadrons and mesons and $K(T,m)$ is a function that is only determined by the mass of the $i$th particle and the temperature. $B_i,Q_i,S_i$ are baryon, charge and strange numbers which are integers and $\hat{\mu}_x=\mu_x/T$.
\begin{figure}[htbp]
  \centering
  \includegraphics[width=0.42\textwidth]{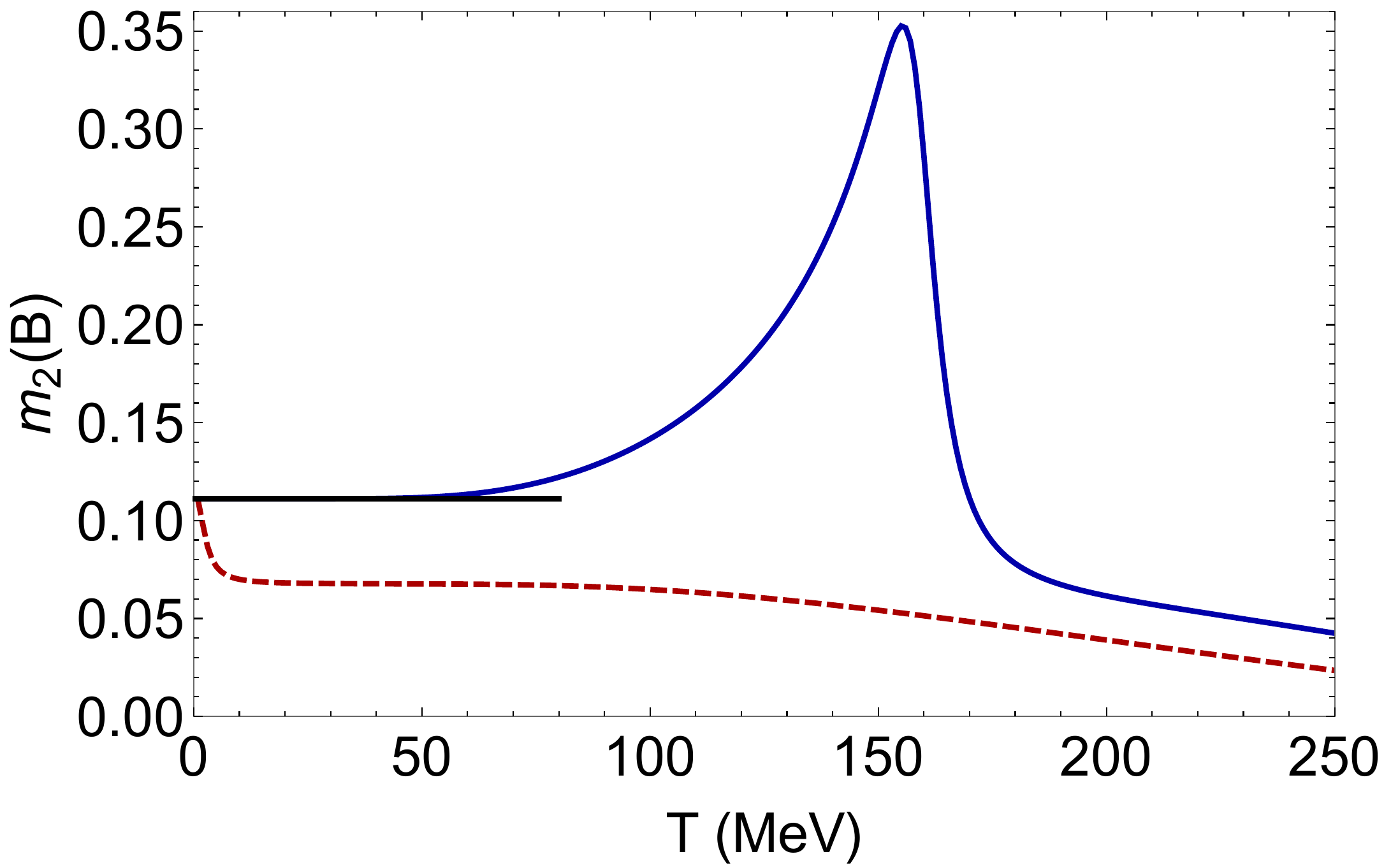}\\
  \caption{Blue solid line: $m_2(B)$ calculated by the NJL model, red dashed line: $m_2(B)$ from a free quark gas model at zero chemical potential. They both tend to the $1/9$ limit at low temperature}\label{fig:m2B_at_zero_chem_potential}
\end{figure}

However, when it comes to the NJL model at quark level, the situation is a little bit different. We still has $m_1(B)=0$ at low temperature and zero chemical potential, but we have $m_2(B)=1/9$ now. This is due to the fact that NJL model does not have confinement. We have $m_2(i)=T^2\chi_i^{(4)}/\chi_i^{(2)}=1 \ (i=u,d,s)$ for the susceptibilities diagonal in flavor, whereas the non-diagonal quark number susceptibilities are much smaller than the diagonal ones because we identify the three quark chemical potentials as independent variables so the change of one quark chemical potential only affects another quarks' number density by modifying its mass, which is much smaller than the direct $\mu$ derivative of $H(m,\mu)$ at low temperature (for example, $\frac{\partial H(m_i,\mu_i)}{\partial\mu_j}\approx\chi_{i,i} \delta_{i,j}$). So we have (also see the definition of $\chi_B^{(n)}$ in Eqs.~\eqref{equ:susbasechange}):
\begin{equation}\label{}
\begin{aligned}
 & m_2(B)=T^2\chi_B^{(4)}/\chi_B^{(2)}=\\
 &\frac{T^2\frac{1}{81}(\chi_u^{(4)}+\chi_d^{(4)}+\chi_s^{(4)})}{\frac{1}{9}(\chi_u^{(2)}+\chi_d^{(2)}+\chi_s^{(2)})}=\frac{1}{9}
\end{aligned}
\end{equation}
This is also observed in Fig.~17 of Ref.~\cite{chen2016robust} in which $m_2(B)$ tends to a value close to $0$ but not $1$ as $m_1(B)$ approaches $0$. However in Ref.~\cite{fukushima2008phase} which adopted a PNJL model, they calculated a ratio $m_2(q)=T^2\chi_q^{(4)}/\chi_q^{(2)}=9$ (meaning $m_2(B)=1$) at low temperature. This is due to the $e^{-3(E\pm\mu)/T}$ factor in the PNJL thermodynamic potential at low temperature which can be viewed as three quarks moving together to form a hadron. Nevertheless, for studying the qualitative behavior of the susceptibilities, the NJL model description is still a good choice, and we will adopt the PNJL model or Dyson-Schwinger equation (DSE) method in the future hoping to obtain better quantitative agreement with experiment \cite{fukushima2008phase,Zhao:2014oha,Wang:2015tia,Xu:2015vna,Cui:2016zqp}.

\section{Summary}\label{sec:summary}
We have studied the fluctuations of conserved charges, i.e., the baryon number, the electric charge number and the strangeness, using a modified $3$ flavor Nambu-Jona-Lasinio model at finite temperature $T$ and baryon chemical potential $\mu_B$. With a simple variation of the four-point coupling inspired by the OPE method, the quark condensate at finite temperature and zero chemical potential of lattice result and our model calculation are in good correspondence. So it is necessary to further test this model by calculating the susceptibilities at finite chemical potential. Same qualitative features are observed as in previous work \cite{chen2016robust}. By using freeze-out curves fitted to experiment data, we studied the energy dependence of conserved quantities fluctuations, and found that the baryon number fluctuations from the NJL model show a similar non-monotonic energy dependence trend as observed in the net-proton fluctuations measured by the STAR experiment. This consistency may indicate that the intriguing structure observed in the experimental data is due to criticality. But this needs more careful studies with more realistic simulation of the dynamics of the heavy-ion collisions and the physics of critical behavior. The effect of smaller $\mu_s$ was discussed and found to be minor in the present study. The discrepancy of $m_2(B)$ between the NJL model, HRG model and experiment result at zero chemical potential can give a sign of quark confinement. Measuring $m_1(B)$ and $m_2(B)$ at collision energy smaller than a few $GeV$ is a good way to test the correctness of the NJL model prediction at large chemical potential. So future experimental measurements of baryon fluctuations at even lower energies are of great interest.

\acknowledgments

The work is supported in part by the MoST of China 973-Project No. 2015CB856901; and the National Natural Science Foundation of China (under Grants No. 11575069, 11221504,No. 11275097, No. 11475085, and No. 11535005).

%


\bibliographystyle{apsrev4-1}
\bibliography{sus}

\end{document}